\documentclass[11pt]{article}
\pdfoutput=1
\usepackage{graphicx,amssymb,amsfonts,amsmath,amssymb,amscd,amstext, mathrsfs,color}
\makeatletter \@addtoreset{equation}{section} \makeatother

\newcommand{\be}{\begin{eqnarray}}
\newcommand{\ee}{\end{eqnarray}}
\newcommand{\ba}{\begin{array}}
\newcommand{\ea}{\end{array}}
\newcommand{\bal}{\begin{align*}}
\newcommand{\eal}{\end{align*}}

\newcommand{\nn}{\nonumber}

\renewcommand{\(}{\Big(}
\renewcommand{\)}{\Big)}
\renewcommand{\[}{\Big[}
\renewcommand{\]}{\Big]}
\def \<{\langle}
\def \>{\rangle}
\definecolor{ggg}{rgb}{0,.6,0}
\begin{document}
~
\vspace{0.5cm}
\begin{center} {\Large \bf  Second-order Perturbative  OTOC of  Anharmonic  Oscillators }
\\
                                            
\vspace{1cm}

Wung-Hong Huang*\\
\vspace{0.5cm}
Department of Physics, National Cheng Kung University,\\
No.1, University Road, Tainan 701, Taiwan\\
                      
\end{center}
\vspace{1cm}
\begin{center} {\large \bf  Abstract} \end{center}
The out-of-time-order correlator  (OTOC) of   simple harmonic  oscillator with extra anharmonic (quartic) interaction  are calculated  by the second quantization method.   We obtain the analytic formulas  of spectrum, Fock space states and matrix elements of coordinate to the second order of anharmonic interaction. These relations clearly reveal the property that the correction of the interaction is proportional to the quantum number of the energy level, and shows the enhancement for some physical quantities. The analytic results explicitly show that   the OTOC is  raising  in the quadratic power law  at early times. Then, we  use the formulas to do numerical  summation to calculate the OTOC, which shows that, at late times, while the first-order perturbative OTOC is oscillating as that in a simple harmonic oscillator,   the second-order perturbative  OTOC    saturates to a constant value.  We compare it with  $2\langle x^2\rangle_T\langle p^2\rangle_T$, which is associated with quantum chaotic behavior in systems that exhibit chaos, and  discuss  the validity of the second-order perturbation.
\\
\\
\\
\scalebox{0.65}{\hspace{0.75cm}\includegraphics{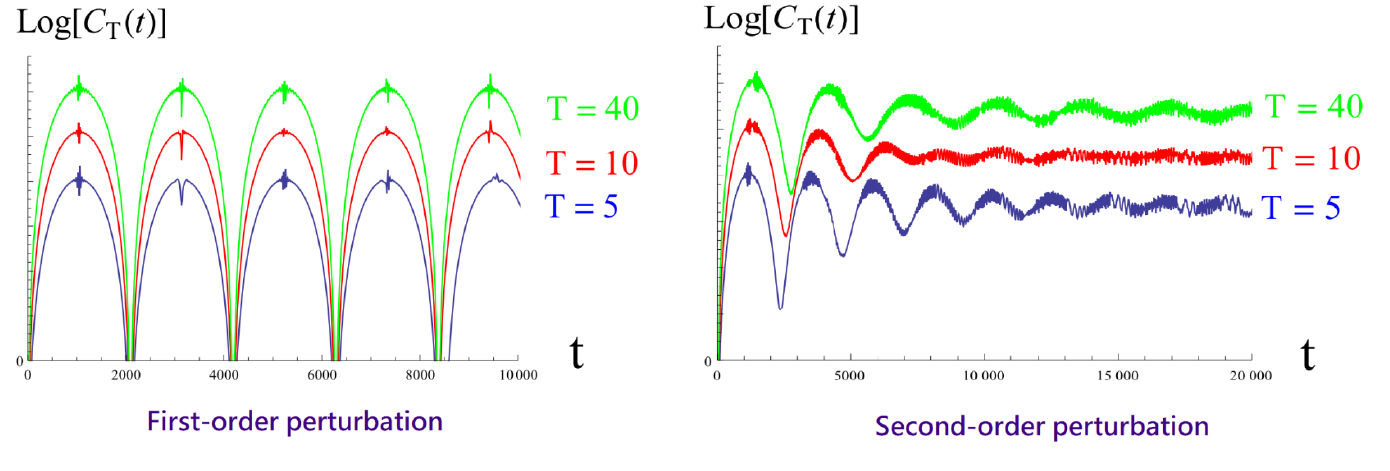}}
\\
\begin{flushleft}
* Retired Professor of NCKU, Taiwan. \\
* E-mail: whhwung@mail.ncku.edu.tw
\end{flushleft}
\newpage
\tableofcontents
\section{Introduction}
Many years ago Larkin and Ovchinnikov introduced the out-of-order correlator (OTOC) to calculate vertex corrections for superconductor currents  \cite{Larkin}.  The function OTOC is defined by
\be
 C_T(t)=-\<[W(t),V(0)]^2\>_T
\ee
where the operators $W$ and $V$ are in the Heisenberg picture, i.e. $W(t)=e^{iHt/\hbar}W(0)e^{-iHt/\hbar}$ by using the Hamiltonian $H $ of the system and  the notation  $\< O \>_T=$Tr $(e^{-\beta H}O)/$ Tr$e^{-\beta H}$  denotes the thermal expectation value at temperature T.    Recently Kitaev \cite{Kitaev15a,Kitaev15b}  revived the quantity and  then, it was  found that the Sachdev-Ye-Kitaev (SYK) model \cite{Kitaev15a,Kitaev15b, Sachdev} develops the exponential growth in thermal OTOC \cite {Maldacena15,Maldacena16, Kitaev15c} 

  The  discover  that   the quantum Lyapunov exponent which associated to the chaos  saturates the bound \cite {Maldacena15} have attracted  a lot of attention in the physics community across many different fields, including quantum information, conformal field theory, AdS/CFT duality \cite{Shenker13a, Shenker14, Roberts14,  Shenker13b,Susskind,Liam,Verlinde,Kristan}.    The relation  between the OTOC and chaos can be easily seen in the semiclassical limit if  W(t) = x(t)  and V = p.   In the classical-quantum correspondence the  commutation relation  is replaced by Poisson bracket : $[A, B]/i\hbar\to \{A,B\}$   and 
\be
-{1\over\hbar^2}\<[x(t),p(0)]^2\to \{x(t),p(0)\}^2=\({\partial x(t)\over \partial x(0) }{\partial p\over\partial p }-{\partial x(t)\over\partial p  }{\partial p\over \partial x(0)}\)^2= \({\partial x(t)\over \partial x(0)}\)^2 
\ee
thus we find that OTOC  $C(t)  =\hbar^2({\partial x(t)\over \partial x(0)})^2 $.  It is known that the classically chaotic system has a Lyapunov exponent $\lambda$ with ${\partial x(t)\over \partial x(0)}\sim  e^{\lambda t}$ which gives  the sensitivity to initial conditions, the classical diagnostic of the butterfly effect \cite{  Maldacena16,Roberts14}.  In this way the quantum OTOC should grow as $\sim  e^{2\lambda t}$ and we can read off the quantum Lyapunov exponent $\lambda$  from OTOC.

The general  OTOC between a pair of local operators $W(t,x)$, $V(0 )$ can be expanded as   
\be
 C_T(t)
&=& \<W(t)V(0)V(0)W(t)\>_T+ \<V(0)W(t)W(t)V(0) \>_T   \nn\\
&&  -  \<W(t)V(0)W(t)V(0)\>_T-\<V(0)W(t)V(0)W(t)\>_T 
\ee
The  first 2 terms are not directly sensitive to chaos  \cite{Shenker14}. For sufficiently late times, they are simply the time-independent disconnected diagram $\(\<W(t) W(t)\>_T\)\cdot\(\<V(0)V(0)\>_T\)$. The  last two terms involves operators with unusual time ordering, hence the name “OTOC”, they are genuine out-of-time-order correlators  and has an universal behavior  
\be
 & &{\<V(0)W(t)V(0)W(t)\>_T\over \<V(0)V(0))\>_T\<W(t)W(t)\>_T}=1-e^{\lambda_L(t-t_*-{|x|\over v_B})} 
\ee
in which
\\
$\bullet$ $t_*$ is a time scale called the scrambling time at which the commutator grows to be ${\cal O}(1)$. 
\\
$\bullet$ $v_B$ is the  buttery velocity   which characterizes speed at which the perturbation grows.  
\\
$\bullet$  $\lambda_L$  is the  Lyapunov exponent   which measures the rate of growth of chaos .
\\

The three parameters have been studied and calculated from gravity side in which some interesting properties, such as  the Lyapunov exponent is bounded by temperature T : 
\be
\lambda \le  {2\pi  T}
\ee
 was found \cite{Maldacena16}. Along this way several literature  had studied the problem under external field or higher gravity theory, for example in \cite{Andrade,Sircar, Kundu,Ross,Huang16,Huang17,Huang18}.

The method of quantum theory to calculate OTOC  with general Hamiltonian  was set up by Hashimoto recently in  \cite{Hashimoto17,Hashimoto20a,Hashimoto20b}.  For  simple  harmonic oscillator (SHO) the OTOC can be calculated exactly and  is purely oscillatory.  Some  complicated examples, such as the two-dimensional stadium billiard \cite{Hashimoto17,Rozenbaum2019}, the Dicke model \cite{Chavez-Carlos2018}, bipartite systems \cite{Prakash2020,Prakash2019} could  exhibit classical chaos in which, after  numerical calculation shows that OTOCs   are growing exponentially at early times followed by a saturation at late times.  In an interesting paper \cite{Hashimoto20a}, the early-time exponential growth of OTOCs, which is expected to exhibit quantum chaos, was found in a system of non-linearly coupled oscillators, which  is reminiscent of gauge theory \cite{Savvidy}.  The method has also been applied to study several systems  including many-body physics, for example in \cite{Das, Romatschke,Shen,Swingle-a, Cotler, Rozenbaum,Dymarsky,Bhattacharyya, Morita, HLi, Lin,Sundar,Swingle}.

In this framework  the properties of  OTOC  are mostly found by using numerical methods. In a  previous unpublished note  \cite{Huang2013} we begin to study OTOC by analytic method under perturbation in second  quantization method.  The quantization method has an advantage that we can find the property of quantum level for any number ``n" while the wavefunction approach could find the property of quantum level only for finite number ``n"  after numerical  evaluations step by step.  It is hoped that the method is able to see how the analytic  results from perturbative calculation could reveal the properties found in numerical calculation, at least qualitatively.   However, it is fail as we does not see the exponential growth in the initial time nor the saturation to a constant OTOC in the final time.

As the previous note  considers  the perturbation  only to first order in coupling strength one may wonder that if the higher-order perturbation could show properties related to the quantum chaos system.  In this  paper we   will first investigate the  second-order perturbative  OTOC of  traditional harmonic  oscillator of quadratic potential while with extra anharmonic (quartic) interaction  in second  quantization method\footnote{The reference  \cite{Romatschke} studied the OTOC of   oscillators with pure  quartic interaction in wavefunction approach.}.  It is interesting that we could use the obtained analytic formula to see that the OTOC will rapidly raise in the quadratic power law  at early times. Especially,  after numerically thermal average we also find that OTOC  saturates to a constant value at later times 
\be
C_T(\infty)\to 2\<x^2\>_T\<p^2\>_T
\ee
 which is known to associate with quantum chaotic behavior in systems that exhibit chaos \cite{Maldacena16}.

This paper is organized as follows. In Section 2, we first review  Hashimoto's method  of  calculating quantum mechanic  OTOC  in SHO and then, use the second quantization method to obtain the same result quickly.

 In Section 3, we  use the   second quantization method to calculate OTOC in the systems  of  harmonic  oscillator with extra anharmonic (quartic) interation.  We obtain analytic formulas  of spectrum, Fock space states and matrix elements of coordinate to the second order of anharmonic interaction.  We  analyze  the  OTOC at early times and show that it is raising in the quadratic power law  at early times.

 In Section 4,   using the analytic formulas formulas in section III we perform numerical summation to get OTOC and compare the results with those in \cite{Romatschke}.   

 In Section 5,  we  analyze  the  OTOC  late  times, and compare it with $2\<x^2\>_T \<p^2\>_T$. We discuss the validity of the second-order approximation of this paper. 

Last section  is devoted to the short discussions and mention some future studies.  
\section{OTOC in Quantum Theory }
\subsection{Quantum Mechanic Approach to OTOC}
We first review the  Hashimoto's method of calculating  OTOC in quantum mechanic  model \cite{
 Hashimoto17}.     For a general time-independent Hamiltonian: $H = H(x_1,....x_n,p_1,....p_n)$ the function of  OTOC is defined by
\be
C_T(t)=-\<[x(t),p(0)]^2 \>_T  
\ee
where $\<{\cal O)}\>\ \equiv tr(e^{-\beta H}\,{\cal O)})/tr e^{-\beta H}$.  Using energy eigenstates $|n\>$,  defined by
\be
H|n\>=E_n|n\>~~~~~~~~~\label{E}
\ee
 as the basis of the Hilbert space, then
\be
  \boxed{C_T(t)={1\over Z}\sum_ne^{-\beta E_n}\,c_n(t)} ,~~~c_n(t)\equiv -\<n|  [x(t),p(0)]^2         |n\>    \label{TC}
\ee
To calculate $c_n(t)$ we can insert  the  complete set $\sum_m|m\>\<m|=1$ to find an useful formula
\be
 \boxed{ c_n(t)}  & =& -\sum_m\<n|  [x(t),p(0)]|m\>\<m|  [x(t),p(0)] |n\>\  \boxed{ =   \sum_m(ib_{nm})(ib_{nm})^*} ~\label{cn}\\
b_{nm}&=& -i\<n|  [x(t),p(0)]|m\>,~~b_{nm}^*=b_{mn} 
\ee
To proceed, we substituting a relation $ x(t) = e^{ iHt/\hbar  }\,x\,e^{- i  Ht/\hbar} $ and inserting the completeness relation again, and finally  obtain
\be
 b_{nm}&\equiv& -i\<n|  x(t), p(0)|m\> +i\<n| p(0) x(t),|m\>=-i\<n|  e^{ i  Ht /\hbar}\,x\,e^{- i  Ht/\hbar} p(0)|m\> +i\<n| p(0) e^{ i  Ht /\hbar}\,x\,e^{- i  Ht/\hbar}|m\>\nn\\
&=&-i\sum_k\<n|  e^{ i\  Ht /\hbar}\,x\,e^{- i Ht/\hbar}|k\>\<k| p(0)|m\>+i\sum_k<n| p(0) |k\>\<k|e^{ i  Ht /\hbar}\,x\,e^{- i  Ht/\hbar}|m\>  \nn\\
&=& -i\sum_k\(e^{i  E_{nk}t/\hbar}x_{nk}p_{km}-e^{i E_{km}t/\hbar}p_{nk}x_{km}\)
\ee
where
\be
E_{nm}= E_n-E_m,~~~ x_{nm}=\<n| x|m\> ,~~p_{nm}=\<n|p |m\>~~~~\label{Enm}
\ee
\\
For a  Hamiltonian\footnote{{\color{blue}Notice that Hashimoto \cite{Hashimoto17} used $H=\sum_i{p_i^2}+U(x_1,....x_N)$ which is that in our notation for M=1/2.  Therefore the formula $b_{mn}$ in eq.(\ref{bnm})  becomes Hashimoto's   formula if M=1/2 and $\hbar=1$.}}
\be
&&H=\sum_i{p_i^2\over 2M}+U(x_1,....x_N)~~\to~~[H,x_i] = - i\hbar { p_i\over M} 
\ee
where  $M$ is the particle mass. Using the relations
\be
p_{km}&=&\<k|p |m\>={iM\over \hbar }\<k|[H,x] |m\>={iM\over \hbar }\<k|(H\,x)-(x\,H) |m\>\nn\\
&=&{iM\over \hbar }\<k|(E_k\,x)-(x\,E_m) |m\>={iM\over \hbar }(E_{km})x_{km} 
\ee
and we have a simple formula
\be
\boxed{ b_{nm} = { M\over \hbar }\sum_k\,x_{nk}x_{km}\(e^{i  E_{nk}t/\hbar} E_{km} -e^{i  E_{km}t/\hbar}E_{nk}\)} ~~~~~~~\label{bnm}
\ee
Now  we can compute OTOC through (\ref{bnm}) once we know $x_{nm}$ and $E_{nm}$ defined in (\ref{Enm}).

In the below, we first summarize the Hashimoto's wavefunction method of  calculating  OTOC for SHO and then, present our method of second quantization method  to obtain the same result. 
\subsection{Simple Harmonic oscillator :   Wavefunction Method}
 For SHO model, which is an integrable example and no chaos, the Hamiltonian $H$, spectrum $E_n$ and state wavefunction $\Psi_n(x)$ have been shown in any textbook of quantum mechanics. 
\be
&& H={p^2\over 2M}+{M \omega^2\over2}x^2,~~~E_n=\hbar\omega\left(n+\frac{1}{2}\right),~~~E_{nm}=\hbar\omega(n-m)  \\
&&\Psi_n(x) = {1\over \sqrt{2^n\,n!}}\({M\omega\over \pi\hbar}   \)^{1/4}\,e^{-{M\omega x^2\over 2\hbar}}\,H_n\(\sqrt{M\omega\over \hbar}\,x\)~~~n =0,1,2,\cdots   
\ee
The functions $H_n$ are the physicists' Hermite polynomials.  Using the orthogonal and  recurrence relations of Hermite polynomials 
\be
\int dx H_m(x)H_n(x)e^{-x^2}&=&\sqrt \pi \ 2^n\ n! \ \delta_{nm}\\
H_{n+1}(x)&=&2x H_{n}(x)-2nH_{n-1}(x) 
\ee
we can   easily calculate $x_{nm}$.  The result  is  
\be
  x_{nm}&=&\int_{-\infty}^{\infty} dx \,\Psi^*_n(x)\,x\,\Psi_m(x)\\
&=& {1\over \sqrt{2^n\,n!}}{1\over \sqrt{2^m\,m!}}\({1\over \pi}   \)^{1/2}\,\int_{-\infty}^{\infty} dx e^{-{M\omega x^2\over  \hbar}}  H_n\(\sqrt{M\omega\over \hbar}\,x\)\ \sqrt{M\omega\over \hbar}\,x \     H_m\(\sqrt{M\omega\over \hbar}\,x\) \nn \\
 &=&{1\over \sqrt{2^n\,n!}}{1\over \sqrt{2^m\,m!}}\({1\over \pi}   \)^{1/2}\,\int_{-\infty}^{\infty} dx e^{-{M\omega x^2\over  \hbar}}  H_n\(\sqrt{M\omega\over \hbar}\,x\)\ \nn \\
 &&~~~~~~~~~~~~~~~~~~~~~~~~~~~ ~~~~~\times   \[ {1\over2} H_{m+1}\(\sqrt{M\omega\over \hbar}\,x\) +mH_{m-1}\(\sqrt{M\omega\over \hbar}\,x\)\]  \nn  \\      
 &=&{1\over \sqrt{2^n\,n!}}{1\over \sqrt{2^m\,m!}}\({1\over \pi}   \)^{1/2}\,\sqrt{\hbar\over M\omega}\, \int_{-\infty}^{\infty} dy e^{-y^2 }  H_n(y) \[ {1\over2} H_{m+1}(y)+mH_{m-1}(y)\]  \nn  \\     
&=& \sqrt{\hbar\over  2M\omega}\, \(\sqrt{n }  \ \delta_{n,m+1}+\sqrt{n+1 }  \ \delta_{n,m-1}\), ~~~~~~n,m=0,1,2,\cdots~~~~~~\label{xnm}
  \ee
Substituting above expressions into  (\ref{Enm}) and  (\ref{bnm}) we obtain 
\be 
 b_{nm}(t)&=&{ M\over \hbar }\sum_k\,x_{nk}x_{km}\(e^{i  E_{nk}t/\hbar} E_{km} -e^{i E_{km}t/\hbar}E_{nk}\) \nn \\
& {=}& \hbar  \( - {   n \cos (  \omega t)} +  {  (n+1) \cos (  \omega t )}   \)\ \delta_{m,n}         \label{first}         \\
&=&   \hbar \cos (\omega t)\ \delta_{nm}
\ee
The first term in (\ref{first}) is coming from $k=n-1$ while second term is from $k=n+1$. Then
\be
c_n(t)&=&   \hbar^2  \cos^2 (  \omega t),~~~C_T(t)=  \hbar^2 \cos^2(  \omega t) ~~~~~~~~~~~\label{HR}
\ee
It is seen that both of $c_T(t)$ and $C_T(t)$ are periodic functions.  They do not depend on energy level $n$ nor temperature $T$. This is a special property   for the harmonic oscillator among  several examples.
\subsection{Simple Harmonic oscillator :  Second Quantization Method}
In the second quantization the states denoted as $|n\>$ are   created and destroyed by the operatrors $a^\dag$ and $a$ respectively. There are following basic properties
\be
&&[a,a^\dag]=1,~~~[a,a]= [a^\dag,a^\dag]=0,~~~~\<n|m\>=\delta_{m,n}\\
&&a^\dag|n\>= \sqrt{n+1} |n+1\>,~~~  a|n\>= \sqrt{n} |n-1\>,~~~a^\dag a |n\>=n |n \>
\ee
Applying above  relations and following definitions
\be
&&x=\sqrt{\hbar\over2M\omega}(a^\dag+a ),~~~p= i\sqrt{M\omega\hbar\over2}(a^\dag-a) 
\ee
to SHO, then, the Hamiltonian $H$, spectrum $E_n$ and $E_{nm}$ become 
\be
  H&=&{p^2\over 2M}+{M\omega^2\over2}x^2=-{\omega\hbar\over4 }(a- a^\dag)^2+{\omega\hbar\over4 }(a+ a^\dag)^2\nn\\
 &=&{\hbar\omega\over2 } \(   a a^\dag +   a^\dag a   \)   = \hbar\omega \( a^\dag a+ \frac{1}{2}\) \\
H|n\>&=&E_n|n\>,~~E_n=\hbar\omega\left(n+\frac{1}{2}\right), ~~~E_{nm}=\hbar\omega(n-m) 
\ee
Basic  relations
\be
x|n\>=\sqrt{\hbar\over2M\omega}(a^\dag+a ) |n\>=\sqrt{\hbar\over2M\omega}\ \sqrt{n}|n-1\>+  \sqrt{\hbar\over2M\omega}\ \sqrt{n+1} |n+1\>
\ee
quickly leads to
\be
x_{nm}&\equiv&\<n|x|m\>=\sqrt{\hbar\over2M\omega}\( \sqrt{m}\ \delta_{n,m-1}+   \sqrt{m+1}\ \delta_{n, m+1}\)   \label{SQxnm}
\ee
which exactly reproduces (\ref {xnm}) and, therefore the values of  $b_{nm}$, $c_n(t)$ and $C_T(t)$ in (\ref{HR}).   Note that the extending to the  supersymmetric quantum  harmonic oscillator was studied in \cite{Das}.
%
\section{Perturbative OTOC of Anharmonic Oscillator : Basic Analytical Formulas }
Wavefunction approach to OTOC of anharmonic oscillator (AHO) with pure  quartic interation had been studied in \cite{Romatschke} following the Hashimoto paper \cite{Hashimoto17}. In this paper we will analyze the OTOC of  standard simple harmonic  oscillator while with extra anharmonic (quartic) interaction  by the  second quantization method in perturbation approximation.  Note that first-order calculations have been introduced in our previously unpublished note \cite{Huang2013}. The second-order calculations in this paper are relatively complex while can show some new interesting  properties.
\subsection{Perturbative Energy and State : Formulas}
We consider the anharmonic oscillators with quartic interaction  
\be
 H&=&({p^2\over 2M}+{M\omega^2\over 2}x^2)+g{x^4}\nn\\
&=&  \hbar\omega \( a^\dag a+ \frac{1}{2}\) +{g } \ \({\hbar\over2M\omega}\)^2(a^\dag+a )^4=H^{(0)} +gV
\ee
which has  a well-known unperturbed solution
\be
H^{(0)} |n^{(0)} \>=E^{(0)}_n |n^{(0)} \>=\hbar\omega\(n^{(0)} +{1\over2}\) |n^{(0)}  \> 
\ee
The second-order perturbation formulas in quantum mechanics textbook are 
\be
{E_n}&=& E^{(0)}_n +g \ \<n^{(0)} | V   |n^{(0)}  \> + g^2 \sum_{k\ne m}{| \<k^{(0)} |V|n^{(0)} \> |^2   \over E^{(0)}_n -E^{(0)}_k}+ {\cal O}(g^3)\\
  {|n\>}&=& |n^{(0)} \>+g \sum_{k\ne n}|k^{(0)} \>{\<k^{(0)} |  V    |n^{(0)} \> \over E^{(0)}_n -E^{(0)}_k }  -{1\over2}g^2\  |n^{(0)}  \>\sum_{k\ne n}{| \<k^{(0)} |V|n^{(0)} \> |^2   \over (E^{(0)}_n -E^{(0)}_k)^2}     \nn   \\
&& -g^2 \sum_{k\ne n}|k^{(0)} \>{\<k^{(0)} |  V    |n^{(0)} \>  \<n^{(0)} |  V    |n^{(0)} \> \over (E^{(0)}_n -E^{(0)}_k )^2}+  g^2 \sum_{k\ne n}\sum_{\ell\ne n}|k^{(0)} \>{\<k^{(0)} |  V    |\ell^{(0)} \> \over E^{(0)}_n -E^{(0)}_k }{\<\ell^{(0)} |  V    |n^{(0)} \> \over E^{(0)}_n -E^{(0)}_\ell } \nn \\
&&+ {\cal O}(g^3)
\ee
In the rest of this section we will use   above formulas to calculate: 

1.  Perturbative Energy   $\widetilde E$. 

2. Perturbative state $\widetilde{|n\>}$. 

3. Perturbative matrix elements $\widetilde{x_{mn}}$. 
\\
 Using these analytic results we can  explicitly show that  the OTOC is  raising  in quadratic power law form at early times. Then, we use the formulas to do numerical  summation to calculate the OTOC  in the next section.
\subsection{Perturbative Energy $\widetilde E$  : Model Calculations}
Using above formulas we can calculate the  corrections to state and energy by extra potential $V$. From now on we  change the notation $\{ n^{(0)}, E^{(0)}  , | n^{(0)}\> \}\to \{ n , E   , | n \> \}$ and   notations $ \{  \widetilde n , \ \widetilde E,  \ \widetilde{|   n   \>}  \} $ are used   to represent the perturbed quantities.  We use the unit : $\hbar=\omega=M=1$ and keep the coupling strength $g$ as a only free parameter.

A crucial quantity we need, after calculation, is
\be
\<k|V|n\>&=& \frac{  \sqrt{n-3} \sqrt{n-2} \sqrt{n-1} \sqrt{n} }{4 }\ \delta _{k,n-4}+\frac{  \sqrt{n-1} \sqrt{n}\ (2 n-1)   }{2  }\ \delta _{k,n-2} +\frac{ 3 (2 n (n+1)+1)   }{4 }\ \delta _{k,n }\nn\\
&&+\frac{  \sqrt{n+1} \sqrt{n+2}\ (2 n+3)   }{2 }\ \delta _{k,n+2}  +\frac{  \sqrt{n+1} \sqrt{n+2} \sqrt{n+3} \sqrt{n+4}   }{4  }  \ \delta _{k,n+4}
\ee
Using above result the perturbative   energy $\widetilde {E_n}$  becomes
\be
\widetilde {E_n} &=& \(n+{1\over2}\) +\frac{3 g   (2 n (n+1)+1) }{4 }-{g^2 (2 n+1) (17 n (n+1)+21)\over8} +{\cal O}(g^3) ~~\label{NE} \\
 \widetilde {E_{nm}} & =&\widetilde {E_n} -\widetilde {E_m}
\ee
This relation shows an interesting  property of   “enhancement" which we mention  it in below.

In the case of  $n> 1$ above relation leads to 
\be
\widetilde {E_n}  &\approx &  \alpha_0  n+ \alpha_1 g   n^2+ \alpha_2 g^2  n^3=n\( \alpha_0 + \alpha_1\cdot ( g   n) + \alpha_2 \cdot(g n)^2\)~~~~~\label{crucial}
\ee
and we see that, for example,  for small value of  $g=0.01$  the value of $gn$ is larger then $1$  if the mode of  state $n>100$. This leads to a general property of  “enhancement"  that no matter how small the value of coupling strength $g$, some perturbative quantities could be very large  in higher energy levels. Therefore, to make the perturbation reliable, we should keep only low-energy-level systems.

 This then limits us to consider systems at low energy temperatures, since  the higher energy level states will be   suppressed by the Boltzmann factor. The problem of valid of perturbation will be discussed in a later section. 

\subsection{Perturbative  State $\widetilde{|n\>}$  and  Matrix Elements $\widetilde{x_{mn}}$ : Model Calculations}
To proceed we have to calculate the perturbative  state $\widetilde {|n\>}$ which  is more complex and is expressed as 
\be
\widetilde {|n\>}&=& |n\> +g   |n\>_{g}+g^2   |n\>_{g^2}+  {\cal O}(g^3) ~~~\label{Nn}
\ee
where
\be
|n\>_{g}&=&\frac{1}{16} \sqrt{n-3} \sqrt{n-2} \sqrt{n-1} \sqrt{n} \ |n-4\>+\frac{1}{4} \sqrt{n-1} \sqrt{n} (2 n-1)\ |n-2\>\nn\\
&&-\frac{1}{4} \sqrt{n+1} \sqrt{n+2} (2 n+3)  \ |n+2\>-\frac{1}{16} \sqrt{n+1} \sqrt{n+2} \sqrt{n+3} \sqrt{n+4} \ |n+4\>~~~
\ee
and
\be
|n\>_{g^2}&=&\frac{1}{512}   \sqrt{n-7} \sqrt{n-6} \sqrt{n-5} \sqrt{n-4} \sqrt{n-3} \sqrt{n-2} \sqrt{n-1} \sqrt{n}\ |n-8\>\nn\\
&&  +\frac{1}{192}   \sqrt{n-5} \sqrt{n-4} \sqrt{n-3} \sqrt{n-2} \sqrt{n-1} \sqrt{n} (6 n-11)
 |n-6\>   \nn\\
&&+\frac{1}{16}   \sqrt{n-3} \sqrt{n-2} (n-1)^{3/2} \sqrt{n} (2 n-7)  \ |n-4\>    \nn\\
&&+\frac{1}{64}   \sqrt{n-1} \sqrt{n} (n (n (2 n+129)-107)+66)\ |n-2\>  \nn\\
&&+\frac{1}{256}   (n (n+1) (65 n (n+1)+422)+156)  \ |n \>\nn\\ 
&& +\frac{1}{64}   \sqrt{n+1} \sqrt{n+2} (n (n (123-2 n)+359)+300)  \ |n+2\>    \nn\\
&& +\frac{1}{16}   \sqrt{n+1} (n+2)^{3/2} \sqrt{n+3} \sqrt{n+4} (2 n+9) \ |n+4\>          \nn\\  
&& +\frac{1}{192}   \sqrt{n+1} \sqrt{n+2} \sqrt{n+3} \sqrt{n+4} \sqrt{n+5} \sqrt{n+6} (6 n+17) \ |n+6\>          \nn\\
&& +\frac{1}{512}   \sqrt{n+1} \sqrt{n+2} \sqrt{n+3} \sqrt{n+4} \sqrt{n+5} \sqrt{n+6} \sqrt{n+7} \sqrt{n+8}  \ |n+8\>      
\ee
After inverse we can find an useful  relation
\be
 |n\>&=&\widetilde {|n\>}-g   |n\>_{g}-g^2   |n\>_{g^2}+  {\cal O}(g^3)                                   \\
&=&\frac{1}{512} g^2 \sqrt{n-7} \sqrt{n-6} \sqrt{n-5} \sqrt{n-4} \sqrt{n-3} \sqrt{n-2} \sqrt{n-1} \sqrt{n}          \  \widetilde  { |n-8\> }    \nn\\
& &+\frac{1}{192} g^2 \sqrt{n-5} \sqrt{n-4} \sqrt{n-3} \sqrt{n-2} \sqrt{n-1} \sqrt{n} (6 n-19) \   \widetilde  { |n-6\>}   \nn\\
& &+\frac{1}{16} g \sqrt{n-3} \sqrt{n-2} \sqrt{n-1} \sqrt{n} \left(g \left(2 n^2-3 n-2\right)-1\right) \ \widetilde  { |n-4\>}     \nn\\
&&  -\frac{1}{64} g \sqrt{n-1} \sqrt{n} \left(g \left(2 n^3-135 n^2+157 n-90\right)+32 n-16\right) \    \widetilde  { |n-2\> }   \nn\\
& &+  \frac{1}{256} \left(256-g^2 \left(65 n^4+130 n^3+487 n^2+422 n+156\right)\right) \  \widetilde  { | n\>}    \nn\\
& &  -\frac{1}{64} g \sqrt{n+1} \sqrt{n+2} \left(g \left(2 n^3+141 n^2+433 n+384\right)-16 (2 n+3)\right) \  \widetilde  { |n+2\> }    \nn\\
& &+ \frac{1}{16} g \sqrt{n+1} \sqrt{n+2} \sqrt{n+3} \sqrt{n+4} \left(g \left(2 n^2+7 n+3\right)+1\right)                            \widetilde  { \   |n+4\> }    \nn\\
& &+\frac{1}{192} g^2 \sqrt{n+1} \sqrt{n+2} \sqrt{n+3} \sqrt{n+4} \sqrt{n+5} \sqrt{n+6} (6 n+25) \   \widetilde  {|n+6\> }     \nn\\
& &+\frac{1}{512} g^2 \sqrt{n+1} \sqrt{n+2} \sqrt{n+3} \sqrt{n+4} \sqrt{n+5} \sqrt{n+6} \sqrt{n+7} \sqrt{n+8}      \                       \widetilde  { |n+8\> }   
\ee
Using above relations the matrix elements  $\widetilde  {x_{mn}} $ could be easy to calculate. The results are 
\be
\widetilde  {x_{mn}} &\equiv& \widetilde{\<m|} x\widetilde{|  n   \>}=\widetilde{\<m|}{1\over \sqrt2}(a^\dag+a)\widetilde{|  n   \>} \\
&=&\frac{g^2 \sqrt{n+1} \sqrt{n+2} \sqrt{n+3} \sqrt{n+4} \sqrt{n+5}  }{16 \sqrt{2}}  \       \delta_{m,n+5}                      \nn\\
&&+\frac{g \sqrt{n+1} \sqrt{n+2} \sqrt{n+3} (4-39 g (n+2)) }{16 \sqrt{2}} \  \delta_{m,n+3}                        \nn\\
&&+\frac{\sqrt{n+1} \left(g^2 \left(303 n^2+606 n+378\right)-48 g (n+1)+32\right)  }{32 \sqrt{2}}  \   \delta_{m,n+1}                            \nn\\
&&+\frac{\sqrt{n} \left(g^2 \left(303 n^2+75\right)-48 g n+32\right)  }{32 \sqrt{2}}    \   \delta_{m,n-1}                         \nn\\
&&+ \frac{g \sqrt{n-2} \sqrt{n-1} \sqrt{n} (4-39 g (n-1))  }{16 \sqrt{2}}  \   \delta_{m,n-3}                    \nn\\
&&+   \frac{g^2 \sqrt{n-4} \sqrt{n-3} \sqrt{n-2} \sqrt{n-1} \sqrt{n} }{16 \sqrt{2}} \  \delta_{m,n-5}                   ~~\label{Nxmn}                        
\ee
In the case of  $n>1$   we can write a general relation
\be
\widetilde  {x_{mn}}  &\approx &   \sqrt n\ \sum_i\( \alpha_{i0} + \alpha_{i1}\cdot  (g   n) + \alpha_{i2}\cdot (g n)^2\)\ \delta_{n,m+i}~~~~~\label{crucial2}
\ee
Compare above relation to  $\widetilde  {E_{m }} $  in (\ref{crucial})   we see that the property of   “enhancement"  also shows in the  matrix elements  $\widetilde  {x_{mn}} $.
\subsection{Perturbative  OTOC at Early Times : Quadratic Power Law}
It is interesting to see that, at early times, the second-order OTOC has a simple analysis form.   First, denote formula $b_{nm}$ in  (\ref{bnm}) in perturbative state by
\be 
 \widetilde{b_{nm}} &=& \sum_k\ \widetilde{x}_{nk}\widetilde{x}_{km}\(\widetilde{E}_{km}  e^{i\widetilde{E} _{nk}t}  -\widetilde{E}_{nk}e^{i\widetilde{E}_{km}t}\)   \label{Nbnm} 
\ee
in the unit $M=\hbar=1$ we can   substitute the analytic form  of $\widetilde{x_{mn}}$ in (\ref{Nxmn}) and  the analytic form  of $\widetilde{E_{mn}}$ in (\ref{NE}) into it  to calculate the associated  microcanonical OTOC formula $c_{n}(t)$ in  (\ref{cn})  in the perturbative state. Then, use the series expansion formulas of the function  $\widetilde{b_{nm}}$  in Appendix A  we can find that
\be
\widetilde{b_{nm}}&=&- \frac{3}{2}  g^2 \sqrt{n-3} \sqrt{n-2} \sqrt{n-1} \sqrt{n}\ t^2 \ \delta_{m,n-4}       \nn \\
&&+     \frac{3}{4} g \sqrt{n-3} \sqrt{n}   (5 g (2 n-1)-4)    \ t^2   \   \delta_{m,n-2}     \nn       \\
&&+    \frac{1}{2} \left(  (6 g (g (6 n (n+1)+3)-2 n-1)-1)+2\right)   \ t^2   \   \delta_{m,n }      \nn      \\
&&+    \frac{3}{4} g \sqrt{n-4} \sqrt{n+1}   (5 g (2 n+3)-4)    \ t^2    \   \delta_{m,n+2}    \nn    \\
&&-     \frac{3}{2}   g^2 \sqrt{n-4} \sqrt{n+1} \sqrt{n+2} \sqrt{n+3}\ t^2     \   \delta_{m,n+4}    +{\cal O}(g^3)+{\cal O}(t^4)    \label{5.1}
\ee
Then, function $c_n(t)$ becomes
\be
c_n(t)&=& 1+  \left(18 g^2 \left(2 n^2+2 n+1\right)-6 g (2 n+1)-1\right) t^2 +{\cal O}(g^3)+{\cal O}(t^4)  \label{5.2}
\ee
which, after thermal average,  implies that
\be
\text{Log}[C_T(t)]\sim \alpha(g)\cdot t^2+{\cal O}(t^3)
\ee
It is a  quadratic power law function, not the exponential law shown in chaos system. This property was also observed in previous literature \cite{Romatschke}, which consider pure quartic potential,  through numerical calculation in wavefunction approach.

\section{Perturbative OTOC of Anharmonic Oscillator : Numerical Summations}
  In this section we will use the analytic formulas presented in Appendix A to do numerical calculation to see more properties of   the OTOC   as the function of time. 
\subsection{First-order Perturbative OTOC of Anharmonic Oscillator}
In the first-order approximation  the analytic form of product $\widetilde{x}_{nk}\widetilde{x}_{km} \widetilde{E}_{km}$ and  product $\widetilde{x}_{nk}\widetilde{x}_{km} \widetilde{E}_{nk}$ in (\ref{Nbnm}) shall be first expressed in leading order of  potential strength $g$. Also, the function $ \widetilde{E}_{nk}$ in factor $e^{i \widetilde{E}_{nk}t }$ and function $ \widetilde{E}_{km}$ in factor $e^{i \widetilde{E}_{km}t }$ are also expressed in leading order of  potential strength $g$.

 Then,   we  calculate $  \widetilde{b_{nm}} $  after the summation over intermediate states   $k$. The analytic form can expressed as
\be
 \widetilde{b_{nm}}& =&\gamma_1  \delta_{m,n+4}+\gamma_2 \delta_{m,n+2}+\gamma_3 \delta_{m,n}+\gamma_4  \delta_{m,n-4}+\gamma_5 \delta_{m,n-4}\\
\gamma_3&=& \cos ( 3 g (n+1)t+t )-n \cos (3 g n t+t))
\ee 
Other components (including $ g^2$ order) are more complex and are represented in Appendix A. 

Using the  explicitly form of $  \widetilde{b_{nm}} $ we then calculate the microcanonical OTOC   $ {c}_n(t)$  by  formula
\be
 {c}_n(t)&=&\sum_m(i\widetilde{b}_{nm})(i\widetilde{b}_{nm})^*                    \label{F}
\ee
We plot  figure 1  to see how the  microcanonical OTOC evolutes with time. The coupling strength is  $g=0.001$  in here.
\\
\\
\scalebox{0.55}{\hspace{7cm}\includegraphics{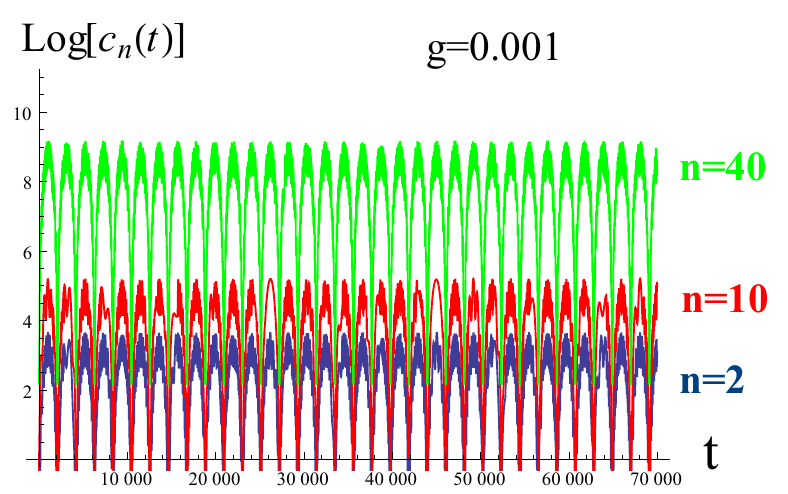}}
\\
{Figure 1:  First-order microcanonical OTOCs  $c_n(t)$, described in (\ref{F}),  as the function of time. The parameter  is chosen to be  $g=0.001$ and $n=2,10,40$.}
\\

The thermal OTOC, $C_T (t)$, is calculated by formula (2.3). We plot figure 2   to see how the
thermal OTOC evolves over time.
\\
\\
\scalebox{0.55}{\hspace{7cm}\includegraphics{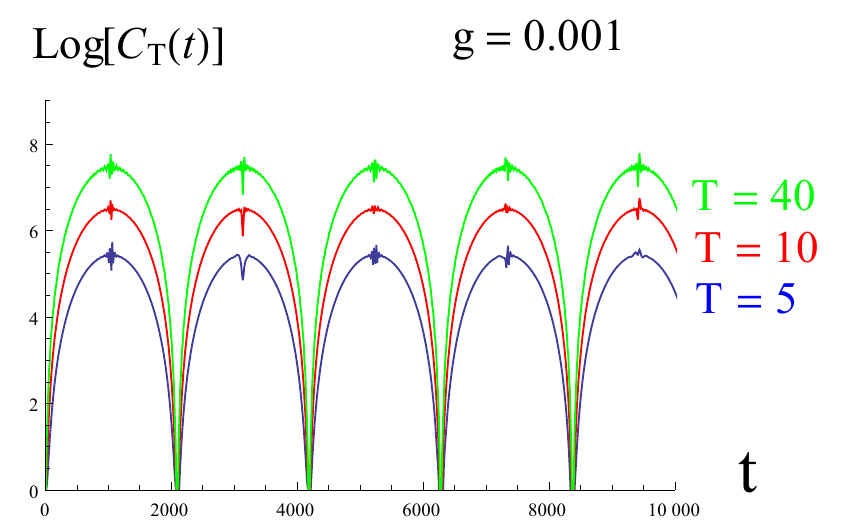}}
\\
{Figure 2:  First-order thermal OTOC   $ {c}_n(t)$, described in (\ref{F}),  as the function of time. The parameter  is chosen to be  $g=0.001$ and $T=5,10,40$.}
\\

Let us make three comments about above result :

1.  The figure 1 and  figure 2 show  that the  OTOC  is an   oscillation function without saturate to a constant value. However, the saturation property was observed in a previous literature  {\cite{Romatschke}, to characterize the similar quantum chaos in the  anharmonic oscillators. This leads us to the second-order calculations in the next section.

2. Note that the microcanonical OTOC  of a free oscillator shown in (\ref{HR}) is $ {c}_n(t) =\cos^2(t)$ which is always $\le 1$. Then one can ask a question : Why a small parameter $g=0.001$ could modify the result to such a large derivation, as shown in figure 1 and figure 2 ? One of the reason is the  “enhancement" property  mentioned below eq.(\ref{crucial2}).  This property tells us that the relative magnitude of   matrix element  $\widetilde  {x_{mn}}$ between zero-order and first-order  perturbation could be not small at large energy level state.

3.  In fact, there is another  “enhancement" mechanism. From eq.(\ref{Nxmn}) we see that at zero order there are only 2 state contributions, (those from $\delta_{m,n+1}$ and $ \delta_{m,n-1}$). While  at first  order there are   4 state contributions,  and at second order there are   6 state contributions.  Therefore in calculating $\widetilde  {b_{nm}}(t)$ the factor  from $\widetilde  {x_{xn}} \widetilde  {x_{km}} \widetilde  {E}_{km}$  tells us that there are    $((2\times2),(4\times4),(6\times6))$  states in each order of $( g^0,g^1,g^2 )$. And in calculating microcanonical OTOT $c_n(t) \sim \widetilde  {b_{nm}}\widetilde  {b^*_{nm}}$  there are   $((2\times2)^2,(4\times4)^2,(6\times6)^2)$  states in each order of "g".  Totally there are (16, 236, 1296) states  in each order of $( g^0,g^1,g^2 )$.  This also gives another  property  of   “enhancement"  while increasing the order of perturbation.

In next subsection we will see that, to second-order perturbation, the  “enhancement" property could dramatically  modify the late stage behave of OTOC and shows the  saturation property   in  the quantum chaos phase.
\subsection{Second-order Perturbative OTOC of Anharmonic Oscillator}
In the second-order approximation  the analytic forms of   $\widetilde{x}_{nk}\widetilde{x}_{km} \widetilde{E}_{km}$,    $\widetilde{x}_{nk}\widetilde{x}_{km} \widetilde{E}_{nk}$ in (\ref{Nbnm}) and the function $ \widetilde{E}_{nk}$ in factor $e^{i \widetilde{E}_{nk}t }$  are all  expressed in second order of  coupling constant $g$.  Then,   we  calculate $  \widetilde{b_{nm}} $ by   summing  the intermediate states   $k$. The analytic form can expressed as
\be
 \widetilde{b_{nm}}& =&\alpha_1  \delta_{m,n+6}+\alpha_2  \delta_{m,n+4}+\alpha_3 \delta_{m,n+2}+\alpha_4 \delta_{m,n}+\alpha_5  \delta_{m,n-4}+\alpha_6 \delta_{m,n-4}+\alpha_7 \delta_{m,n-6}~~~~~\label{A}\\
\alpha_1&=&-\frac{1}{32} \(5 e^{\frac{1}{4} i t (3 g^2  (17 n^2+34 n+24 )-12 g (n+1)-4)}  -5 e^ {\frac{1}{4} i t  (3 g^2  (17 n^2+204 n+619 )-12 g (n+6)-4 )  }\nn\\
&&+ 3 e^{  \frac{3}{4} i t \left(g^2 \left(51 n^2+204 n+259\right)-12 g (n+2)-4\right) }+e^{  \frac{5}{4} i t \left(g^2 \left(51 n^2+306 n+582\right)-12 g (n+3)-4\right) } \nn\\
&&-e^{\frac{5}{4} i t \left(g^2 \left(51 n^2+408 n+939\right)-12 g (n+4)-4\right)}-3 e^{\frac{3}{4} i t \left(g^2 \left(51 n^2+510 n+1330\right)-12 g (n+5)-4\right) }\)   \nn\\
&&\times g^2 \sqrt{n+1} \sqrt{n+2} \sqrt{n+3} \sqrt{n+4} \sqrt{n+5} \sqrt{n+6} 
\ee 
Other components are more complex and are represented in Appendix A. 

Using the  explicitly form of $  \widetilde{b_{nm}} $ we then calculate the microcanonical OTOC   $ {c}_n(t)
$ by formula  (\ref{F}).   After numerically summations we plot  figure 3  to see how the  microcanonical OTOC  evolves over time : 
\\
\scalebox{0.55}{\hspace{6cm}\includegraphics{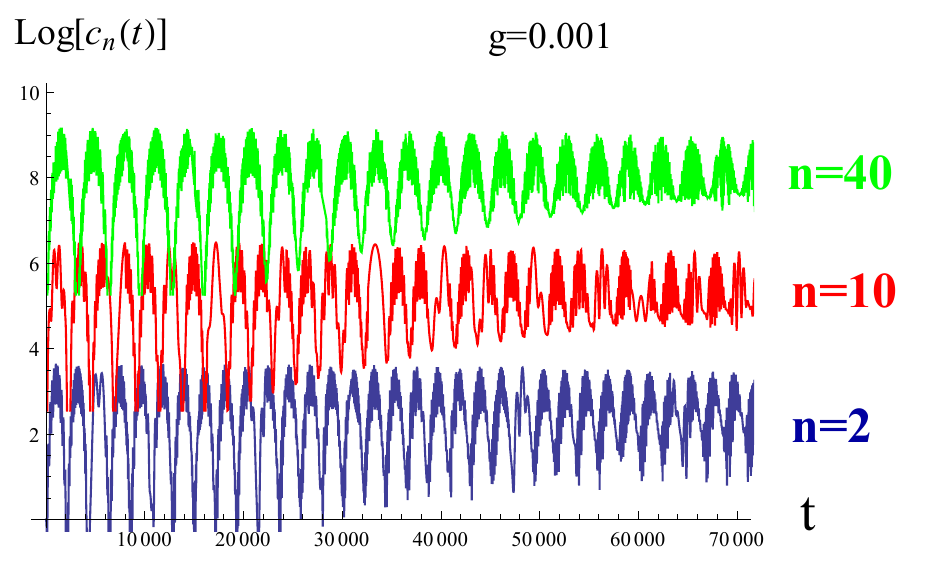}}
\\
{Figure 3:  Second-order microcanonical OTOC  as the function of time. The parameter  is chosen to be  $g=0.001$ and energy level $n=2,10,40$.}
\\
\\
The property of saturation is shown in figure 3 for the second second-order microcanonical OTOC,   contrasts  to the  figure 1 for the first-order microcanonical OTOC.

The thermal  OTOC, $ C_T(t)$, is calculated by formula  (\ref{TC}).  We plot figure 4 in below to see how the  thermal OTOC  evolves over time : 
\\
\\
\scalebox{0.5}{\hspace{ 6.5 cm}\includegraphics{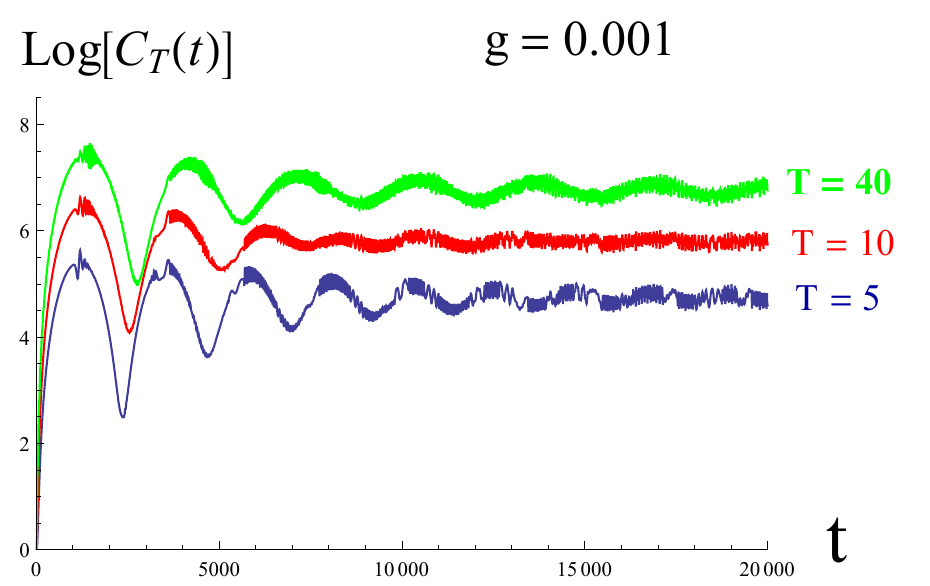}}
\\
{Figure 4:  Second-order thermal OTOC  as the function of time. The parameter is chosen to be   g=0.001    for temperature  T=5,10,40  respectively.
\\

Note that the precise value of thermal OTOC in the perturbation will depend on the number of state we adopt in the statistical average. This figure is the result of summation  from n=1 to n=40, i.e. $n_{\text{cut}}=40$. The problem of choosing value of cutoff level $n_{\text{cut}}$ is discussed later.

 We see that, in later stage, the second-order perturbative OTOC  shows the  saturation property as that found in systems that exhibit  quantum chaos phase. This property was observed in previous literature \cite{Romatschke} which investigated the system with pure  anharmonic (quartic) interaction by the wavefunction approach. Figure 4  also shows that the stronger the coupling, the faster it saturates to a constant value of OTOC.

\section{OTOC at  Late  Times and Valid of  Perturbation}
The properties of OTOC  at late  times are special and has  useful physical property, we discuss them in this section.
\subsection{Saturation at Late  Times and $2\<x^2\>_T \<p^2\>_T   $}
It is expected that, after the Ehrenfest time,   the thermal OTOC will asymptotically in time \cite{Maldacena16} 
\be
C_T(\infty)\to 2\<x^2\>_T \<p^2\>_T                  \label{xxpp}
\ee
which is associated with quantum chaotic behavior in systems that exhibit chaos \footnote{we denote $ \widetilde{\<x^2\>}$ by  ${\<x^2\>}$ and $ \widetilde{\<p^2\>}$ by  ${\<p^2\>}$  in here for conventional notation}.  

To calculate this quantity we can use the analytical method in section III to derive the analytic forms of $\widetilde{x^2_{nm}}$ and $\widetilde{p^2_{nm}}$. Then we have the following second-order relations
\be
\widetilde{E_n}&=&\frac{1}{8} \left(g^2 (-(2 n+1)) (17 n (n+1)+21)+6 g (2 n (n+1)+1)+8 n+4\right)                 \label{A1} \\
 \widetilde{\<x^2\>}&=& \widetilde{\<n|}  ({1\over \sqrt2}(a^\dag+a))^2  \widetilde{|n\>}        \nn   \\
&=& \frac{1}{512} \left(g^2 (-(2 n+1)) (n (n+1) (73 n (n+1)-1514)-2244)-384 g (2 n (n+1)+1)+256 (2 n+1)\right)    \label{A2}\nn
 \\
\\
 \widetilde{ \<p^2\>}&=& \widetilde{\<n|} ({1\over \sqrt2}(a^\dag-a))^2  \widetilde{|n\>}        \nn   \\
&=&   \frac{1}{512} \left(-3 g^2 (2 n+1) (n (n+1) (19 n (n+1)+786)+852)+384 g (2 n (n+1)+1)+256 (2 n+1)\right)  \label{A3}\nn\\       
\ee
which reduces to $\widetilde{E_n}= \widetilde{\<x^2\>}= \widetilde{ \<p^2\>}=n+{1\over2}$ if g=0 as it shall be. Then, using above relations to do numerical summation to find the associated thermal average. 
The results for the functions  of  thermal  Log[$2\<x^2\>_T \<p^2\>_T]$ and OTOC Log[$C_T(\infty)]$ which is calculated in sec. IV,  as the function of temperature in the case of g=0.001 are shown in figure 5. 
\\
\scalebox{0.6}{\hspace{5 cm}\includegraphics{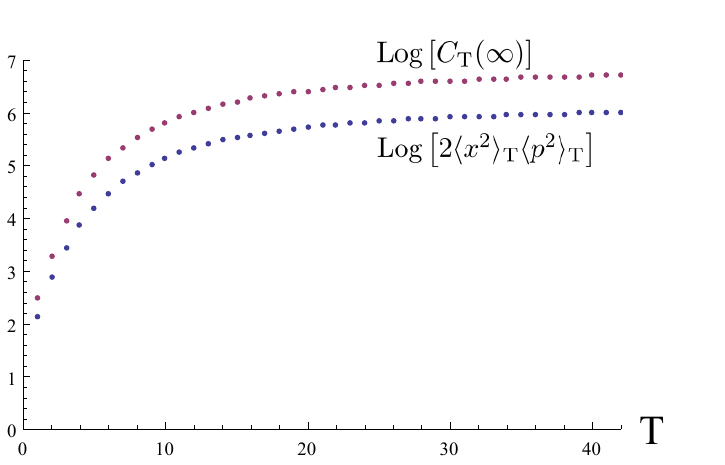}}
\\
{Figure 5 : Thermal  Log[$2\<x^2\>_T \<p^2\>_T]$ and OTOC Log[$C_T(\infty)]$   as the function of temperature. The parameter is chosen to be g=0.001 with  cut-off  value  of  $n_{\text{cut}}$ = 40.}  
\\
\\
 We see that  the asymptotic property of the thermal OTOC and the property of the OTOC in a quantum chaotic system (\ref{xxpp}) are quite similar to each other. Thus we conclude that the saturated parts of the thermal OTOC   are similar to  $2\<x^2\>_T\<p^2\>_T$ in a temperature range,  which is known to associate with quantum chaotic behavior in systems that exhibit chaos \cite{Maldacena16}.   The slight difference between them could come from our truncation of energy levels, i.e.  $n_{\text{cut}}  = 40$. The arguments are as those in the system of coupled harmonic oscillation studied by Hashimoto in  \cite{Hashimoto20a}.  See the figure 6 therein.}
\subsection{Valid of Perturbation}
In this subsection we discuss the problem of the valid of perturbation used in this paper.

 1.  Using the  three analytic equations, (\ref{A2}), and (\ref{A3}), we can see that the three quantities will become negative for large $n$.  In this case the perturbation is invalid  because these three quantities should be  definitively positive.      Consider, for example, the case of g=0.001, which we study in this paper,   if   $n\ge 693$,  $n\ge 43$, $n\ge 47$, the values of $\widetilde{E_n}$,  $\widetilde{\<x^2\>} $,   $ \widetilde{\<p^2\>}$  become negative numbers respectively. Therefore, the system we consider should be subject to the constraint $n<42\equiv n_{\text{cut}}$.

 The dependence of the cutoff value $n_{\text{cut}}$ on the coupling constant $g$ is shown in Figure 6.  We see that $n_{\text{cut}}$ is a decreasing function of $g$. This property is consistent with the  “enhancement"   mechanism mentioned before, in which the perturbation  parameter is $g\cdot n$ instead of $g$ itself.
\\
\\
\scalebox{0.5}{\hspace{5.5 cm}\includegraphics{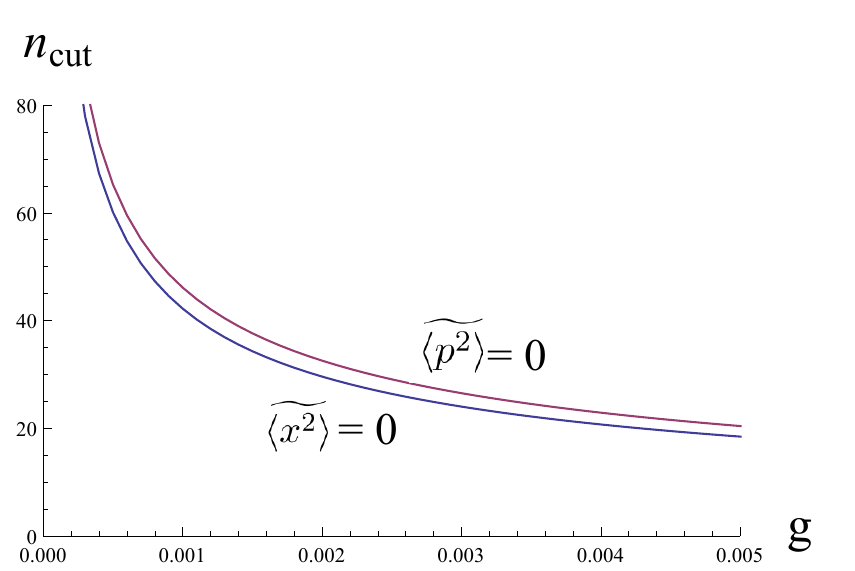}}
\\
{Figure 6 :  The $n_{\text{cut}}$ solutions in equations $\widetilde{\<x^2\>}=0$ and $\widetilde{\<p^2\>}=0$  as  function of coupling $g$. The solutions  are the decreasing functions and  $n_{\text{cut}}$ = 43, 47 respectively for the system with  $g=0.001$.}  
\\

2. The existence of cutoff $n_{\text{cut}}$ in the perturbation method  could also be shown  in the thermal system. For example, we can consider the function  $2\<x^2\>_T \<p^2\>_T$ with various cut-off   $n_{\text{cut}}$. As explicitly  shown in figure 7   that in the case of  $n_{\text{cut}}=55$, if $T>T_d=17.68$ the function becomes a decreasing function of temperature, which is un-physical\footnote{The temperature   $T_d$ may be very high. For example, $T_d\approx 200 $ for $n_{\text{cut}}=47$.}.
\\
\\
\scalebox{0.5}{\hspace{5.5 cm}\includegraphics{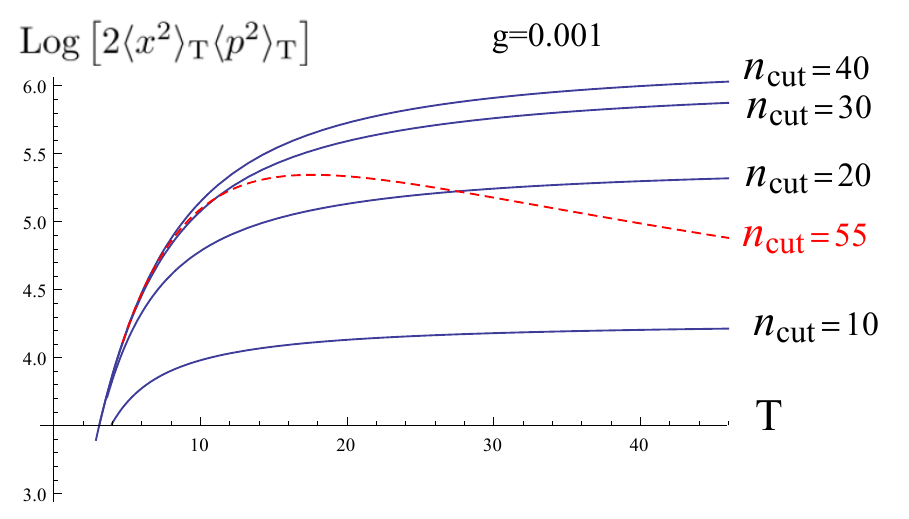}}
\\
{Figure 7 :  Second-order thermal $2\<x^2\>_T \<p^2\>_T$  as the function of temperature. The parameter is chosen to be g=0.001 for various cut-off  values   $n_{\text{cut}}$ = 10, 20, 30, 40, 55.}  
\\

3. It is for these considerations that,   the system we study is with g=0.001 and  the cutoff value in Figure 4 we drew earlier is $n_{\text{cut}}$ = 40. Since the temperature in Figure 4 is $T\le40$, the higher energy levels do not contribute much, so our perturbation results are reliable.
4. To see the precise properties, we draw Figure 8. The figure shows that, for example, for T=1 the levels $ n > 3$ do not contribute much to the thermal average,  for T=5 the levels $ n > 18$ do not contribute much to the thermal average, and so on. Although the figure does not use the property of level of $n>40 $ for $T$ = 10, 20, 40, it shows a nearly flat property. Therefore, we can conclude that the thermal  properties calculated with cutoff $n_{\text{cut}}=40$ could reveal the physical properties.
\\
\\
\scalebox{0.65}{\hspace{5.5 cm}\includegraphics{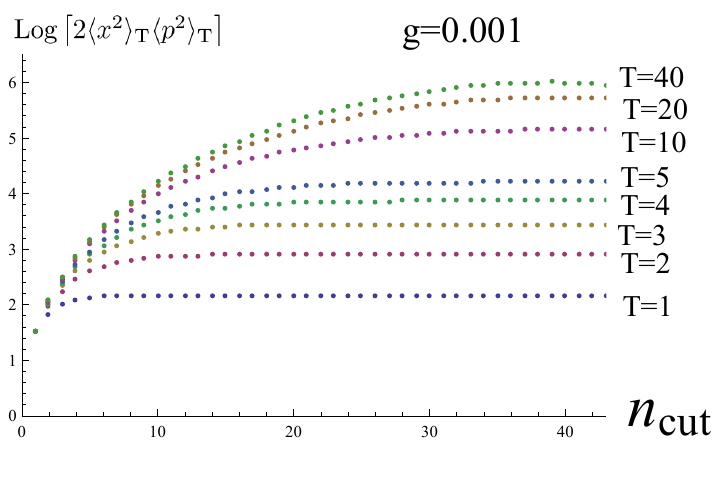}}
\\
{Figure 8 :  Second-order thermal $2\<x^2\>_T \<p^2\>_T$  as the function of cutoff $n_{\text{cut}}$. The parameter is chosen to be g=0.001 with  various cutoff  values $n_{\text{cut}}$ for  each $T$ = 1, 2, 3, 4, 5,10, 20, 40.  The figure shows that levels $ n > 3$ do not contribute much to the thermal average for  for T=1, and so on.}  

4. Finally, follow the above method we can analyze the problem  while with $g=0.01$. The result shows that the cutoff level $n_{\text{cut}}$ and associated temperature T is  $\{T, n_{\text{cut}}\} \approx \{40, 43\} $ in the case of $g=0.001 $ will change to   $\{T, n_{\text{cut}}\} \approx \{4, 12\} $ in the case of $g=0.01 $. This is consistent with the enhanced property: The perturbation parameter is $g\cdot n$. Thus a large $g$ will be paired with a small $n_{\text{cut}}$ and therefore a lower temperature.
 \section{Discusions}
In this paper we use the method which  was set up  by Hashimoto recently in \cite{Hashimoto17,Hashimoto20a,Hashimoto20b} to study the OTOC in the  quantum harmonic oscillator with extra anharmonic (quartic) interaction. In contrast to the previous studies we  calculate  the OTOC  by second quantization method in  perturbative approximation, which can give a few  useful analytical formulas.

Some interesting properties could   be read from the  analytical formulas of perturbative results. For example,  using the analytical formulas we clearly see  that the interactions can enhance the correlation to very large values over time and is proportional to the energy level.  This lead to the rapidly rising characteristic of the initial time of OTOC.  The behavior that at  the early times OTOC  is raising in the quadratic power law is demonstrated without numerical calculations. 

On the other hand, the perturbation method of the second quantization method also has some disadvantages, which we mention below  :

$\bullet$ First, the coupling strength shall be small. Therefore, some properties which appear only in strong coupling cannot be studied. This is the general constrain in the perturbation method.

$\bullet$ Secondly, in the wave function method, the Hamiltonian has to solved step by step to find   the solutions of each energy level $n$. The associated microcanonic OTOC is then calculated. Therefore, we can  know the properties of the solved levels only up to finite $n$. On the other hand, in the second quantization method, we can   know the physical properties of any level $n$, for example $\widetilde {E_n}$, and can find   the associated microcanonic OTOC. However, as discussed in the previous subsection: { Valid of Perturbation}, for each coupling $g$ there is an associated cutoff level $n_{\text{cut}}$ and the computed properties are  reliable  only for the levels $n<n_{\text{cut}}$. So, essentially only the property of   levels up to the finite $n_{\text{cut}}$ are useful.

Finally, we mention two further studies relevant to our approach. 

$\bullet$ First, the OTOC of  non-linearly coupled oscillators was studied in an interesting paper \cite{Hashimoto20a}.   The properties of the early-time exponential growth of OTOC and saturation property at late times, which is shown in systems of quantum chaos, were found.  The first-order perturbation investigation in our unpublished paper  note \cite{Huang2013}, however did not find either of them. It is expected that the  second-order perturbation could exhibit both characteristics.  Note that the coupled oscillator has two coordinate, (x,y), and has two associated quantum level number, ($n_x$,$n_y$), as shown in \cite{Huang2013}. Thus the second-order study of coupled oscillators will be more complicated. However, once we have  a few  useful analytical formulas, then as has shown in this paper, we can find the analytic property at  early times and then  get the Lyapunov  in an efficient way. It is interesting to study the problem. 

$\bullet$ Secondly, the problem of many-body chaos at weak coupling had been investigated in several years ago by Stanford \cite{Stanford}, in which the system of matrix $\Phi^4$ theory was studied. In a recent paper, Kolganov and Trunin study in detail the classical and quantum butterfly effects in related theories \cite{Kolganov}.  Since that the interacting scalar field theory could be transformed to a system of coupled oscillators, see for example \cite{Huang21}, it is interesting to study the problems along the prescription of this paper.
\newpage
\appendix
\section{Function  Form  of $\alpha_i $ in $\widetilde{b_{nm}}$ of eq.(\ref{A})}
\be
 \widetilde{b_{nm}}& =&\alpha_1  \delta_{m,n+6}+\alpha_2  \delta_{m,n+4}+\alpha_3 \delta_{m,n+2}+\alpha_4 \delta_{m,n}+\alpha_5  \delta_{m,n-4}+\alpha_6 \delta_{m,n-4}+\alpha_7 \delta_{m,n-6}~~~~~~~~(\ref{A})\nn
\ee
where 
\be
\alpha_1&=&-\frac{ g^2}{32} \(5 e^{\frac{1}{4} i t (3 g^2  (17 n^2+34 n+24 )-12 g (n+1)-4)}  -5 e^ {\frac{1}{4} i t  (3 g^2  (17 n^2+204 n+619 )-12 g (n+6)-4 )  }\nn\\
&&+ 3 e^{  \frac{3}{4} i t \left(g^2 \left(51 n^2+204 n+259\right)-12 g (n+2)-4\right) }+e^{  \frac{5}{4} i t \left(g^2 \left(51 n^2+306 n+582\right)-12 g (n+3)-4\right) } \nn\\
&&-e^{\frac{5}{4} i t \left(g^2 \left(51 n^2+408 n+939\right)-12 g (n+4)-4\right)}-3 e^{\frac{3}{4} i t \left(g^2 \left(51 n^2+510 n+1330\right)-12 g (n+5)-4\right) }\)\nn \\
&&\times \sqrt{n+1} \sqrt{n+2} \sqrt{n+3} \sqrt{n+4} \sqrt{n+5} \sqrt{n+6} =0 +{\cal O}(g^3)+{\cal O}(t^4)\\
\nn\\
\alpha_2&=& \frac{g}{32} \(-5g ne^{-\frac{1}{4} i t \left(3 g^2 \left(17 n^2+7\right)-12 g n-4\right)}- gne^{\frac{5}{4} i t \left(3 g^2 \left(17 n^2+68 n+109\right)-12 g (n+2)-4\right)}\nn\\
&&+5  (5 + n) g e^{ -\frac{1}{4} i t \left(3 g^2 \left(17 n^2+170 n+432\right)-12 g (n+5)-4\right)} +(5 + n) g e^{ \frac{5}{4} i t \left(g^2 \left(51 n^2+306 n+582\right)-12 g (n+3)-4\right)  } \nn\\
&&+(4-3 g (11 n+37))\ e^{ \frac{3}{4} i t \left(g^2 \left(51 n^2+306 n+514\right)-12 g (n+3)-4\right)} \nn\\
&&+(g (33 n+54)-4) \ e^{\frac{3}{4} i t \left(g^2 \left(51 n^2+204 n+259\right)-12 g (n+2)-4\right)}    \nn\\
&&-3(g (33 n+78)-4)e^{ \frac{1}{4} i t \left(g^2 \left(51 n^2+408 n+837\right)-12 g (n+4)-4\right)  }\nn\\
&&+3(g (33 n+87)-4)\ e^{\frac{1}{4} i t \left(3 g^2 \left(17 n^2+34 n+24\right)-12 g (n+1)-4\right)} \)\times \sqrt{n+1} \sqrt{n+2} \sqrt{n+3} \sqrt{n+4} \\
&=&- \frac{3}{2}  g^2 \sqrt{n-3} \sqrt{n-2} \sqrt{n-1} \sqrt{n}\ t^2  +{\cal O}(g^3)+{\cal O}(t^4)  \\
\nn\\
\alpha_3&=&{1\over 64}\sqrt{n+1}\sqrt{n+2}\ e^{ -\frac{1}{4} i t \left(3 g^2 \left(34 n^2+102 n+167\right)-12 g (2 n+3)-8\right)   }\nn\\
&&\times\(6gn(3 g (11 n+9)-4)e^{ \frac{1}{4} i t \left(g^2 \left(51 n^2+306 n+480\right)-12 g (n+3)-4\right)              }\nn\\
&&+2 g n (g (33 n+39)-4) \ e^{ \frac{1}{4} i t \left(3 g^2 \left(85 n^2+204 n+273\right)-12 g (5 n+6)-20\right) } \nn\\
&&-2 g (n+3) (g (33 n+60)-4)\  e^{  \frac{1}{4} i t \left(3 g^2 \left(85 n^2+306 n+426\right)-12 g (5 n+9)-20\right) }\nn\\
&&-      6 g (n+3) (g (33 n+72)-4)\ e^{\frac{1}{4} i t \left(3 g^2 \left(17 n^2+7\right)-12 g n-4\right)}                             \nn\\
&&+   \left(g^2 \left(-18 n^2+498 n+801\right)-48 g+32\right)  e^{ \frac{3}{4} i t \left(g^2 \left(51 n^2+170 n+242\right)-4 g (3 n+5)-4\right)  }                            \nn\\
&&+   \left(3 g^2 \left(6 n^2+202 n+285\right)-48 g-32\right) \ e^{  \frac{3}{4} i t \left(g^2 \left(51 n^2+136 n+191\right)-4 g (3 n+4)-4\right) }  \)\\
&=&    \frac{3}{4} g \sqrt{n-3} \sqrt{n}   (5 g (2 n-1)-4)    \ t^2    +{\cal O}(g^3)+{\cal O}(t^4) \\
\nn\\
\alpha_4&=&\frac{1}{16} \(n  (9 g^2 (n^2+1 )-16 ) \cos  (\frac{1}{4} t  (3 g^2 (17 n^2+7 )-12 g n-4 ) )  \nn\\
&&-(n+1)  (9 g^2  (n^2+2 n+2 )-16 ) \cos  (\frac{1}{4} t  (3 g^2 (17 n^2+34 n+24 )-12 g (n+1)-4 ) ) \nn\\
&&3 g^2  ( (n^3+6 n^2+11 n+6\ )-n  (n^2-3 n+2 ) \cos  (\frac{3}{4} t  (g^2  (51 n^2-102 n+106 )-12 g (n-1)-4 ) ) \nn\\
&& \cos  (\frac{3}{4} t  (g^2 (51 n^2+204 n+259 )-12 g (n+2)-4 ) ) )\)\\
&=&  \frac{1}{2} \left(  (6 g (g (6 n (n+1)+3)-2 n-1)-1)+2\right)   \ t^2   +{\cal O}(g^3)+{\cal O}(t^4)
\ee
\newpage
\be
 \alpha_5&=& \frac{1}{64} \sqrt{n-1} \sqrt{n} \( (2 g (n-2) (3 g (11 n-9)-4) \ e^{ -  \frac{3}{4}   i t  (g^2  (51 n^2-102 n+106 )-12 g (n-1)-4 ) } \nn\\
&&+6 g (n-2) (g (33 n-39)-4) \ e^ {\frac{1}{4} i t  (3 g^2  (17 n^2-68 n+75 )-12 g (n-2)-4 ) }\nn\\
&&-2 g (n+1) (g (33 n-6)-4)  e^{-\frac{3}{4}  i t  (g^2  (51 n^2+55 )-12 g n-4 ) }\nn\\
&&-6 g (n+1) (g (33 n+6)-4)  e^{\frac{1}{4} i t  (3 g^2  (17 n^2+34 n+24 )-12 g (n+1)-4 )}\nn\\
&&+ (3 g^2  (6 n^2+178 n-95 )-48 g-32 ) \ e ^{-\frac{1}{4} i t  (3 g^2  (17 n^2-34 n+24 )-12 g (n-1)-4 )}\nn\\
&&+ (-3 g^2  (6 n^2-190 n+89 )-48 g+32 ) \ e^{-\frac{1}{4} i t  (3 g^2  (17 n^2+7 )-12 g n-4 ) } \)\\
&=& \frac{3}{4} g \sqrt{n-4} \sqrt{n+1}   (5 g (2 n+3)-4)    \ t^2    +{\cal O}(g^3)+{\cal O}(t^4)      \\
\nn\\
 \alpha_6&=&\frac{g}{32}   \sqrt{n-3} \sqrt{n-2} \sqrt{n-1} \sqrt{n}  \(-g (n-4)  (  e^ { -  \frac{ 5 }{4}  i t  (3 g^2  (17 n^2-68 n+109 )-12 g (n-2)-4 )  }\nn\\
&&-5 g (n-4)  e^ {   \frac{1}{4} i t  (g^2  (51 n^2-408 n+837 )-12 g (n-4)-4 )  }\nn\\
&&+5 g (n+1)  e^ {   \frac{1}{4} i t  (3 g^2  (17 n^2+34 n+24 )-12 g (n+1)-4  )  }\nn\\
&&+g (n+1)  e ^{  - \frac{ 5 }{4}  i t  (3 g^2  (17 n^2-34 n+58 )-12 g (n-1)-4  )  }\nn\\
&&+(g (33 n-78)-4)  e^ {   \frac{-3}{4}   i t  (g^2  (51 n^2-204 n+259 )-12 g (n-2)-4  )  }\nn\\
&&+3 (g (33 n-45)-4)  e^ {   -\frac{1}{4} i t  (g^2  (51 n^2-306 n+480 )-12 g (n-3)-4  )  }\nn\\
&&+(g (21-33 n)+4)  e^ {  \frac{-3}{4}   i t  (g^2  (51 n^2-102 n+106 )-12 g (n-1)-4 )   }\nn\\
&&+(g (162-99 n)+12)  e^ { - \frac{ 1}{4} i t  (3 g^2  (17 n^2+7 )-12 g n-4  )  } \)  \\
&=&-     \frac{3}{2}   g^2 \sqrt{n-4} \sqrt{n+1} \sqrt{n+2} \sqrt{n+3}\ t^2         +{\cal O}(g^3)+{\cal O}(t^4)  \\
\nn\\
 \alpha_7 &=&\frac{ g^2}{32}   \(5  e ^{  -\frac{1}{4} i t  (3 g^2  (17 n^2+7 )-12 g n-4  )  }  -5  e ^{  -\frac{1}{4} i t  (3 g^2  (17 n^2-170 n+432 )-12 g (n-5)-4  )  } \nn\\
 &&+ e ^{ -  \frac{5}{4}   i t  (3 g^2  (17 n^2-68 n+109 )-12 g (n-2)-4 ) } -3  e ^{ -  \frac{3}{4}   i t  (g^2  (51 n^2-408 n+871 )-12 g (n-4)-4 )  } \nn\\ 
&&- e ^{  - \frac{5}{4}   i t  (g^2  (51 n^2-306 n+582 )-12 g (n-3)-4 )   } +3  e ^{ -  \frac{3}{4}   i t  (g^2  (51 n^2-102 n+106 )-12 g (n-1)-4  )  } \)\nn\\
&&\times \sqrt{n-5} \sqrt{n-4} \sqrt{n-3} \sqrt{n-2} \sqrt{n-1} \sqrt{n}= 0+{\cal O}(g^3)+{\cal O}(t^4)
\ee
Expanding the above formulas to the orders of $t^2$ and $g^2$, we get eq.(\ref{5.1}). Adding these terms, we then get eq.(\ref{5.2}), which demonstrates the quadratic power law characteristics of OTOC in the early stages under the perturbation approximation.

\newpage
\begin{center} 
{\bf  \large References}
\end{center}
 

\begin{enumerate}
 \bibitem{Larkin} A. I. Larkin and Y. N. Ovchinnikov,  “Quasiclassical method in the theory of
superconductivity,” JETP 28, 6 (1969) 1200.
 \bibitem{Kitaev15a} A. Kitaev, “A simple model of quantum holography,”  in KITP
Strings Seminar and Entanglement 2015 Program (2015).
 \bibitem{Kitaev15b}A. Kitaev, “Hidden correlations in the Hawking radiation and
thermal noise,” in Proceedings of the KITP (2015).
 \bibitem{Sachdev}  S. Sachdev and J. Ye, “Gapless spin fluid ground state in a random, quantum Heisenberg
magnet,” Phys. Rev. Lett. 70, 3339 (1993) [cond-mat/9212030].
 \bibitem{Maldacena15} J. Maldacena, S. H. Shenker, and D. Stanford, “A bound on chaos,” JHEP 08 (2016) 106 [arXiv:1503.01409 [hep-th]]
\bibitem{Maldacena16}J. Maldacena and D. Stanford, “Remarks on the Sachdev-Ye-Kitaev model,” Phys. Rev. D 94, no. 10, 106002 (2016) [arXiv:1604.07818 [hep-th]].
 \bibitem{Kitaev15c} A. Kitaev and  S. J.  Suh, “The soft mode in the Sachdev-Ye-Kitaev model and its gravity dual,”  JHEP 05 (2018)183 [	arXiv:1711.08467 [hep-th]]


\bibitem{Shenker13a} S. H. Shenker and D. Stanford, “Black holes and the butterfly effect,” JHEP 03 (2014) 067 [arXiv:1306.0622] 
 \bibitem{Shenker14} S. H. Shenker and D. Stanford, “Stringy effects in scrambling,”  JHEP. 05 (2015) 132  [arXiv:1412.6087 [hep-th]] 
 \bibitem{Roberts14} D. A. Roberts and D. Stanford, “Two-dimensional conformal field theory and the butterfly effect,”  PRL. 115 (2015) 131603  [arXiv:1412.5123 [hep-th]] 
\bibitem{Shenker13b} S.H. Shenker and D. Stanford, “Multiple Shocks,” JHEP 12 (2014) 046 [arXiv:1312.3296 [hep-th]]
\bibitem{Susskind} D. A. Roberts, D. Stanford and L. Susskind, ``Localized shocks,'' JHEP 1503 (2015)
051 [arXiv:1409.8180 [hep-th]] 
\bibitem{Liam}  A. L. Fitzpatrick and J. Kaplan, ``A Quantum Correction To Chaos,'' JHEP  05 (2016) 070 [arXiv:1601.06164  [hep-th]] 

 \bibitem{Verlinde} G. J. Turiaci and H. L. Verlinde, ``On CFT and Quantum Chaos,''  JHEP 1612 (2016) 110 [arXiv:1603.03020 [hep-th]]

 \bibitem{Kristan} Kristan Jensen, ``Chaos in AdS2 holography,'' Phys.\ Rev.\ Lett.\    117  (2016) 111601 
  [arXiv: 1605.06098 [hep-th]] 

\bibitem{Andrade} T. Andrade, S. Fischetti, D, Marolf, S. F. Ross and M. Rozali, ``Entanglement and Correlations near Extremality CFTs dual to Reissner-Nordstrom $AdS_5$,'' JHEP 4 (2014) 23 [arXiv:1312.2839 [hep-th]].
\bibitem{Sircar} N. Sircar, J. Sonnenschein and W. Tangarife, “Extending the scope of holographic mutual
information and chaotic behavior,” JHEP 05 (2016) 091 [arXiv:1602.07307 [hep-th]] 
\bibitem{Kundu} S. Kundu and J. F. Pedraza, ``Aspects of Holographic Entanglement at Finite Temperature and Chemical Potential,'' JHEP 08 (2016) 177 [arXiv:1602.07353 [hep-th]] ; 
\bibitem{Ross} A.P. Reynolds and S.F. Ross, “Butterflies with rotation and charge,” Class. Quant. Grav. 33
(2016) 215008 [arXiv:1604.04099 [hep-th]]
\bibitem{Huang16}  Wung-Hong Huang  and  Yi-Hsien Du, “Butterfly Effect and Holographic Mutual Information under External Field and Spatial Noncommutativity,”  JHEP 02(2017)032   [arXiv:1609.08841 [hep-th]] 
\bibitem{Huang17}  Wung-Hong Huang, “Holographic Butterfly Velocities in Brane Geometry and Einstein-Gauss-Bonnet Gravity with Matters,” Phys. Rev. D 97 (2018) 066020  [arXiv:1710.05765 [hep-th]] 
\bibitem{Huang18}  Wung-Hong Huang, “Butterfly Velocity in Quadratic Gravity,” Class. Quantum Grav. 35 (2018)195004  [arXiv:arXiv:1804.05527 [hep-th]] 
\bibitem{Hashimoto17}  K. Hashimoto, K. Murata and R. Yoshii, “Out-of-time-order correlators in quantum
mechanics,” JHEP 1710, 138 (2017) [arXiv:1703.09435 [hep-th]]  
 \bibitem{Hashimoto20a} T. Akutagawa, K. Hashimoto, T. Sasaki, and R. Watanabe, “Out-of-time-
order correlator in coupled harmonic oscillators,” JHEP 08 (2020) 013 [arXiv:2004.04381 [hep-th]]
 \bibitem{Hashimoto20b} K. Hashimoto, K-B Huh, K-Y Kim, and R. Watanabe, “Exponential growth of out-of-time-order correlator without chaos: inverted harmonic oscillator,” JHEP 11 (2020) 068 [
arXiv:2007.04746 [hep-th]]

\bibitem{Rozenbaum2019} 
  E.~B.~Rozenbaum, S.~Ganeshan and V.~Galitski,   ``Universal level statistics of the out-of-time-ordered operator,''   Phys.\ Rev.\ B {\bf 100}, no. 3, 035112 (2019) 
  [arXiv:1801.10591 [cond-mat.dis-nn]].

\bibitem{Chavez-Carlos2018} 
  J.~Ch\'avez-Carlos, B.~L\'opez-Del-Carpio, M.~A.~Bastarrachea-Magnani, P.~Str\'ansk\'y, S.~Lerma-Hern\'andez, L.~F.~Santos and J.~G.~Hirsch,
  ``Quantum and Classical Lyapunov Exponents in Atom-Field Interaction Systems,''
  Phys.\ Rev.\ Lett.\  {\bf 122}, no. 2, 024101 (2019) 
  [arXiv:1807.10292 [cond-mat.stat-mech]].

\bibitem{Prakash2020}
R.~Prakash and A.~Lakshminarayan,
``Scrambling in strongly chaotic weakly coupled bipartite systems: Universality beyond the Ehrenfest timescale,'' Phys. Rev. B \textbf{101} (2020) no.12, 121108
[arXiv:1904.06482 [quant-ph]].

\bibitem{Prakash2019}
R.~Prakash and A.~Lakshminarayan,
``Out-of-time-order correlators in bipartite nonintegrable systems,''
Acta Phys. Polon. A \textbf{136} (2019), 803-810
[arXiv:1911.02829 [quant-ph]].

\bibitem{Savvidy} G. Savvidy, “Classical and Quantum Mechanics of Nonabelian Gauge Fields,” Nucl. Phys. B 246 (1984) 302

 \bibitem{Das} R. N. Das, S. Dutta, and A. Maji, “Generalised out-of-time-order correlator in supersymmetric quantum mechanics,” JHEP 08 (2020) 013 [arXiv:2010.07089 [ quant-ph]]


\bibitem{Romatschke} P.  Romatschke, “Quantum mechanical out-of-time-ordered-correlators for the anharmonic (quartic) oscillator,” JHEP, 2101 (2021) 030  [arXiv:2008.06056 [hep-th]] 

\bibitem{Shen} H. Shen, P. Zhang, R. Fan, and H. Zhai, “Out-of-time-order correlation at a quantum phase
transition,” Phys. Rev. B 96 (2017) 054503 [arXiv:1608.02438 [cond-mat.quant-gas]]

\bibitem{Swingle-a} D. Chowdhury and B. Swingle, “Onset of many-body chaos in the O(N) model,” Phys. Rev D 96 (2017)  065005 [arXiv:1703.02545 [cond-mat.str-el]]

\bibitem{Cotler} J. S. Cotler, D. Ding, and G. R. Penington, “Out-of-time-order Operators and the Butterfly
Effect,” Annals Phys. 396 (2018) 318 [arXiv:1704.02979 [quant-ph]].


\bibitem{Rozenbaum} E. B. Rozenbaum, L. A. Bunimovich, and V. Galitski, “Early-time exponential instabilities in nonchaotic quantum systems,” Phys. Rev. Lett. 125  (2020)  014101 [arXiv:1902.05466].

\bibitem{Dymarsky}  A. Avdoshkin and  A. Dymarsky, “Euclidean operator growth and quantum chaos,” Phys. Rev. Research 2  (2020) 043234 [arXiv:1911.09672].


\bibitem{Bhattacharyya}  A. Bhattacharyya, W. Chemissany, and S. S.  Haque, J. Murugan, and B. Yan, “The Multi-faceted Inverted Harmonic Oscillator: Chaos and Complexity,” SciPost Phys. Core 4 (2021) 002 [arXiv:2007.01232  [hep-th]]

\bibitem{Morita} T. Morita, “Extracting classical Lyapunov exponent from one-dimensional quantum mechanics,” Phys.Rev.D 106 (2022) 106001  [arXiv:2105.09603 [hep-th]] 
\bibitem{HLi}H. Li, E. Halperin, R. R. W. Wang, J. L. Bohn, “Out-of-Time-Order-Correlator for van der Waals potential,” Phys. Rev. A 107 (2023) 032818   	 [arXiv: 2301.10323 [quant-ph]];

\bibitem{Lin} C. J. Lin and O. I. Motrunich, “Out-of-time-ordered correlators in a quantum Ising chain,”   Phys. Rev. B 97 (2018) 144304  [arXiv:1801.01636 [cond-mat.stat-mech]]
\bibitem{Sundar} B. Sundar, A. Elben, L. K. Joshi, T. V. Zache, “Proposal for measuring out-of-time-ordered correlators at finite temperature with coupled spin chains,”  New Journal of Physics, 24 (2022) , 023037 [arXiv:2107.02196 [cond-mat.quant-gas]] 
 \bibitem{Swingle}  S. Xu and B. Swingle, “Scrambling dynamics and out-of-time ordered correlators in quantum many-body systems: a tutorial,” [arXiv: 2202.07060 [hep-th]].
 \bibitem{Huang2013} Wung-Hong Huang, “Perturbative OTOC and Quantum Chaos in Harmonic Oscillators : Second Quantization Method,” [arXiv : 2306.03644 [hep-th]].
\bibitem{Stanford} D. Stanford, “Many-body chaos at weak coupling,”  JHEP 10 (2016) 009  [arXiv:1512.07687 [hep-th]]   
\bibitem{Kolganov} N. Kolganov and D. A. Trunin, ``Classical and quantum butterfly effect in nonlinear vector mechanics,'' Phys. Rev. D 106 (2022) , 025003 [arXiv:2205.05663  [hep-th]].
\bibitem{Huang21} Wung-Hong Huang, “Perturbative complexity of interacting theory,” Phys. Rev. D 103,  (2021) 065002  [arXiv:2008.05944 [hep-th]]  
\end{enumerate}
 \end{document}